# A PROACTIVE FLOW ADMISSION AND RE-ROUTING SCHEME FOR LOAD BALANCING AND MITIGATION OF CONGESTION PROPAGATION IN SDN DATA PLANE

Sminesh C. N.[1], Grace Mary Kanaga E.[2], and Ranjitha K.[3]

[1&3] Dept. of Computer Science and Engineering, Govt. Engineering College, Thrissur, India
[2] Dept. of Computer Science and Engineering, Karunya Institute of Technology and Sciences, Coimbatore, India

***ABSTRACT***

*The centralized architecture in software-defined network (SDN) provides a global view of the underlying network, paving the way for enormous research in the area of SDN traffic engineering (SDN TE). This research focuses on the load balancing aspects of SDN TE, given that the existing reactive methods for data-plane load balancing eventually result in packet loss and proactive schemes for data plane load balancing do not address congestion propagation. In the proposed work, the SDN controller periodically monitors flow level statistics and utilization on each link in the network and over-utilized links that cause network congestion and packet loss are identified as bottleneck links. For load balancing the identified largest flow and further traffic through these bottleneck links are rerouted through the lightly-loaded alternate path. The proposed scheme models a Bayesian Network using the observed port utilization and residual bandwidth to decide whether the newly computed alternate path can handle the new flow load before flow admission which in turn reduces congestion propagation. The simulation results show that when the network traffic increases the proposed method efficiently re-routes the flows and balance the network load which substantially improves the network efficiency and the quality of service (QoS) parameters.*

## 1. INTRODUCTION

SDN architecture decouples the control plane from its data plane, enabling network intelligence to be centralized in SDN controllers. Kreutz et al. [2] mentioned a global, up-to-date view of the underlying network which greatly simplifies network design and the orchestration of a large number of network devices. The SDN controller uses a southbound interface called the OpenFlow protocol to populate and manipulate flow table entries [4] [8]. In [16], OpenFlow allows direct software-based access and manipulation of flow tables that instruct OpenFlow switches in the infrastructure layer to direct network traffic. The challenges in SDN traffic engineering grow when a number of network elements and traffic flows increase. To achieve traffic fairness load balancing, fast failure recovery, and avoid congestion in bursty-traffic scenarios, SDN TE demands dynamic load balancing and fast failure recovery schemes for the data plane which is specified in [1] and [31].

In [28], the existing hash-based equal-cost multi-path (ECMP) routing, load balancing splits the load across multiple paths using flow hashing techniques. The key limitation of this approach is that it does not differentiate between macro and micro flows. The static mapping of paths, without considering network statistics, lead to underutilization of network resources, network congestion and eventual packet loss. The methods Hedera and Mahout proposed in [29] and [30] to treat micro and macro flows differently allow dynamic flow scheduling, but the exhaustive flow statistics monitoring for each flow results in data plane overload.

As a result of inefficiencies in the existing routing architecture, dominantly selected routes are often chosen, culminating in overutilization and congestion of such links, leaving other links idle and resulting unfairness in network load distribution. Park et al. [3] proposed SDN TE solutions that address failure recovery mechanisms for the network element, or link failures and load balancing are either reactive or proactive in nature. In the reactive methods, failure recovery time to establish a restoration path is long, leading to further congestion and packet loss. According to Mohan et al. [15], the proactive approaches result in faster failure recovery; however, the pre-computation of link-node disjoint backup paths and installation of forwarding rules in the additional switches of the backup paths is an overhead. In both approaches above, the selection of alternate paths and backup paths does not consider current network utilization or flow statistics. If the newly computed alternate path or backup path happens to be in a fully utilized link, this re-routing can result in further network congestion in that particular link, leading to congestion propagation.

The proposed method is a proactive traffic engineering scheme for data plane load balancing that analyzes the traffic applying flow level statistics, gathered periodically using OpenFlow messages. The SDN controller periodically monitors the load on each link using OpenFlow port statistics messages. If the utilization of any link exceeds a threshold value, the link concerned is identified as a bottleneck link.

Once bottleneck links are identified, the SDN controller updates the network virtual topology by setting the designated bottleneck link's weight as infinity. The largest flow through these bottleneck links is identified based on byte count from the open flow *flow_stat* structure. To mitigate the possibility of congestion propagation in the network, the residual bandwidth of each link is computed within an observation window. A Bayesian network is modeled using the parameters above to ensure that the newly computed alternate path can accommodate the new flow without leading to congestion propagation. Finally, the alternate route entries are made in the flow table, the identified largest flow and further traffic use this newly computed path. Thus the proposed system, through load balancing, can provide a solution congestion and congestion propagation.

The simulation of experiment topology is done using Mininet and the SDN controller is done using OpenIris controller. The network traffic is generated using D-ITG packet generator as used in [11] and [12]. The proposed system is simulated as three modules.

- The bottleneck link identification module periodically monitors port utilization and identifies links as bottleneck links if the link utilization exceeds a fixed threshold value.
- The alternate route computation module updates the virtual topology, excluding the identified bottleneck links, and computes the shortest alternate path using Dijkstra's algorithm.
- The flow admission module computes the residual bandwidth of each link in the alternate path, and the proposed Bayesian network model decides whether flow admission is possible through the identified alternate path.

A performance evaluation is undertaken to compare the proposed proactive strategy and the existing ARLD reactive re-routing strategy in SDN. It is observed that when the network load increases, the proposed method outperforms the existing reactive strategies in terms of the average packet loss, average throughput and average end-to-end delay.

Section II of this paper explains the concept of SDN architecture and traffic engineering challenges. Section III discusses the proposed proactive strategy for data plane load balancing in SDN. Section IV discusses the performance analysis of the proposed proactive routing algorithms. Section V concludes the paper.

## 2. Review Of Literature

SDN is an emerging network architecture where the control plane logic is decoupled from the forwarding plane and is directly programmable. One can control, change, and manage network behavior dynamically through programming, using open interfaces instead of relying on closed boxes and vendor-defined interfaces as illustrated in [33], [9] and [10]. Network intelligence is logically centralized in SDN controllers which maintain a global view of the network. This centralized up-to-date view makes the controller ideally suited to perform network management functions. The principal function of the data plane layer is packet forwarding and gathering various flow level statistics with minimal intelligence. In [16], the OpenFlow switches communicate with SDN Controller using OpenFlow protocol. OpenFlow protocol uses the concept of flows to identify network traffic based on predefined match rules. These rules can be statically or dynamically programmed by the SDN controller software. The interconnection between OpenFlow switches through OpenFlow ports, the ingress ports receive packets and forwards through output port as given in [34].

SDN is currently accelerating the innovation and evolution of modern data center networks. The unique features of SDN, including global visibility, programmability, and openness, demand highly scalable and intelligent TE techniques. In [7], the authors chiefly focused on traffic engineering methods such as fault tolerance, flow management, topology update, and traffic analysis. C. E. Hopps [28] proposed Switch load-balancing schemes, which are mainly based on the Equal-Cost Multi-Path (ECMP) scheme, split flows across available paths using the flow hashing technique. The ECMP does not differentiate between micro and macro flows; hence two or more large, long-lived flows can collide on their hash and contend for the same output port. To overcome this drawback of the ECMP, two solutions are proposed. In the first solution, Hedera periodically polls the scheduler at the edge switches to collect flow statistics, detects macro flows, dynamically calculates an appropriate path for large flows, and installs these paths on the switches. In the second solution, Mahout monitors and detects macro flows at the end host via a shim layer in the Operating System, instead of directly monitoring the switches in the network. In Mahout, switches are configured to identify macro flows and the Mahout controller then computes a suitable path for it, whereas micro flows are forwarded using the conventional ECMP. Kanagavelu et al. [6] proposed a local re-routing mechanism in SDN, where the OpenFlow controller collects the port, table and flow statistics from all OpenFlow switches at fixed intervals. The routing engine computes the least-loaded, shortest candidate paths between any pair of end hosts, based on these statistics. It checks for congestion periodically across all the links and if any link load exceeds a threshold value, the controller re-routes one or more large flows across the link to an alternate path one by one, ensuring that the large flow will not overload the newly chosen alternate path.

The SDN controller supports the co-existence of multiple routing algorithms. Mohan et al. [15] proposed two proactive local rerouting algorithms for fast failure recovery. The Forward Local Re-routing (FLR) computes a set of backup paths, one for each link in the primary path. The failed traffic is routed in the forward direction from the point of failure to another switch in the downstream of the primary path. If there exist multiple backup paths from the point of failure to a downstream switch on the primary path, the backup path with the least number of additional switches will be selected. The Backward Local Rerouting (BLR) determines only a single backup path for a primary path. For any link failure on the primary path, the failed traffic will be forwarded from the point of failure back to the source and thereafter routed along a link-node-disjoint backup path. The controller pro-actively determines the link-node-disjoint backup path and installs forwarding rules for the additional switches on the backup path. The proactive approaches discussed involve the computation of the primary path and backup path for faster failure recovery, resulting in additional resource reservations and flow table entries as an overhead to the SDN controller.

Owing to the limited visibility of network topology, ensuring load balancing and traffic fairness in the existing network architecture is cumbersome. Sang Min et al. [3] proposed a new routing architecture called Automatic Re-routing with Loss Detection (ARLD), enabled by SDN and OpenFlow protocols. The underlying idea behind the ARLD is that the most common cause in a wired network for packet drop is network congestion, so when a packet drop occurs at a certain link, the controller treats that link as a bottleneck link. Once a packet drop is detected, the SDN controller's re-routing module is invoked, thereby initiating a by-pass or alternate route computation. The re-routing module finds an alternate route, if any, following which the controller update switches flow tables with the alternate route. Further traffic gets routed through this newly added route until the flow table entry expires. The reactive approach discussed in the ARLD provides solutions to failure recovery; however, flow rerouting without considering the current network statistics results in congestion propagation. Kao et al. [18] proposed an effective proactive traffic re-routing mechanism for congestion avoidance using an SDN controller to manage actions and forwarding rules. The controller observes the current traffic of switches and updates the topology according to the weight assigned based on computed bandwidth usage. Then the traffic on the congested link is instantly transferred to available links. This method effectively allocates and utilizes the network bandwidth.

Gholami et al. [19] proposed a method based on SDN for reducing congestion in data center networks. When congestion occurs, the controller computes fixed and variable costs of each link and reroutes traffic through alternate route with minimum load. The simulation results show a significant improvement in throughput and average packet delay reduction when the network load scales up. However being a reactive strategy exchange of OpenFlow messages and computation of the utilization and load of each path is an overhead for the controller.

Hwang et al. [20] proposed a scalable congestion control protocol with an objective to avoid switch buffer overflow and to reduce the queuing delay even under bursty traffic. By monitoring the number of TCP flows that traverse each switch port the fair-share of each flow are computed such that the total link utilization does not surpass the bandwidth delay product. This information eventually passed to each TCP source by updating the advertisement window field in the TCP header. The proposed algorithm transfer the minimum number of flows from the congested link to the backup path hence results in improved QoS and congestion control.

Attarha et al. [21] proposed an algorithm for avoiding congestion, the SDN controller monitor the network status periodically and routes the newly arrived flows through a path which can forward the flow without resulting in congestion. Whenever the utilization exceeds 70% of the link capacity, the controller computes the amount of traffic and shortest backup path through which the flows are to be re-routed to avoid congestion which in turn reduces the load on the congested ports of an over-utilized switch.

As illustrated in [22] and [24], the Bayesian network provides an efficient method to draw conclusions based on a probabilistic approach, since it models a graphical structure that allows us to represent and reason about an uncertain domain with relatively simple visualization. The nodes in a Bayesian network represent a set of random variables, $X = X_1 , ..X_i , ..., X_n$. From the domain, a set of directed arcs connects pairs of nodes, $X_i \rightarrow X_j$, representing the direct dependencies between variables. The strength of the relationship between variables is quantified by the conditional probability distributions associated with each node. The proposed system models a Bayesian network to decide whether flow admission is possible through a path for a dynamically computed Port Utilization and Residual Bandwidth during an observation window as proposed in [17] and [22].

## 3. PROPOSED SYSTEM

From the existing methodologies for SDN congestion control and load balancing, it is observed that to ensure load balancing and traffic fairness in SDN, improved proactive congestion control strategies addressing both traffic re-routing and intelligent flow admission is indispensable. The proposed system analyses the traffic using OpenFlow messages and predicts both congestion and congestion propagation based on the observed parameters.

### 3.1 Problem Definition

To develop a proactive traffic analysis scheme for SDN data plane load balancing using improved flow re-routing and Bayesian Network based flow admission and to evaluate the network performance from the perspective of QoS parameters.

### 3.2 Proposed System Design

The proposed flow monitoring method intends to determine the links which are over-utilized and eventually lead to congestion and packet loss. Such links are identified as bottleneck links before they actually start dropping packets. Using the OpenFlow statistics message, the SDN controller periodically determines port utilization for all the ports in the network. If the utilization of any port increases beyond a threshold value, the corresponding link is identified as a bottleneck link. Once bottleneck links are identified, the SDN controller initiates alternate route computation. After computing the alternate path, if any, the flow admission module models a Bayesian network to decide whether the alternate path can handle the new flow load without leading to congestion propagation. The overall design of the proposed system is given in Fig. 1.

The proposed system is developed in three sub-modules and integrated within the SDN controller. The first module identifies the list of bottleneck links, based on the statistics collected for each port through the OpenFlow port statistics request and reply messages. The second module updates the virtual topology and computes an alternate path. The third module determines the possibility of flow admission using a Bayesian Network model and subsequently updates the flow table entries. The following sections describe the design details and algorithms of the three sub-modules.

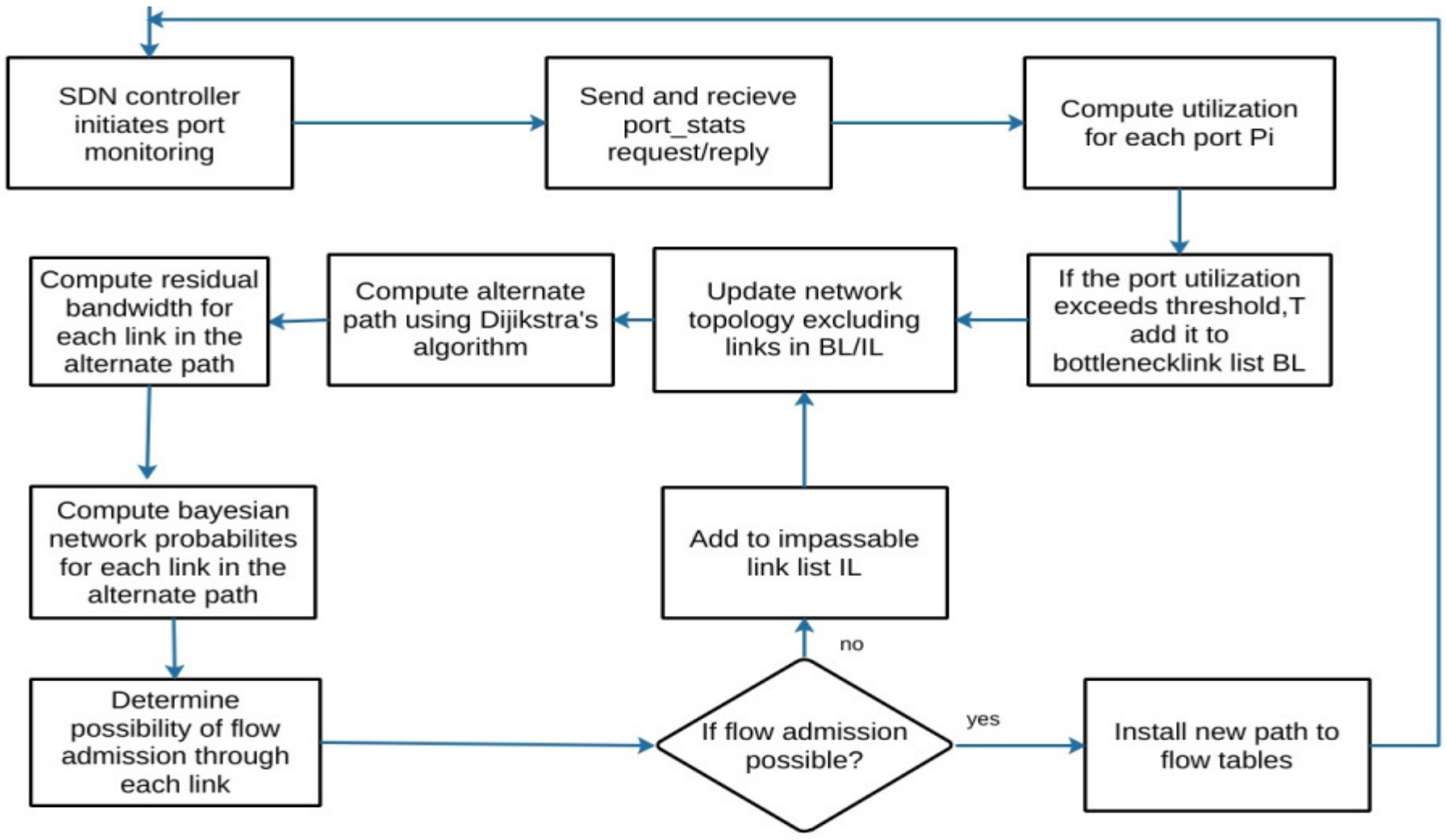

**Fig.1** System Architecture

### 3.2.1 Identifying Bottleneck link

Port monitoring module in the controller performs periodic monitoring of each port on each switch for its utilization. Controller does this by polling each switch every *t=10* seconds for port statistics using Open Flow message *ofp_port_stats*. Controller sends OpenFlow port status request message *ofp_port_stats_request* (with *port_no* = *OFPP_ANY*, for all ports of a switch) to all the switches connected to the controller. Each switch responds by sending *ofp_port_stats_reply* message. Controller then computes port utilisation for each port using transmitted bytes count *tx_bytes* using the formula:

$$PortUtilization = \left( \frac{BytesTransmitted * 8}{PortSpeedTimeInterval} \right) * 100 \tag{1}$$

If the port utilization exceeds the fixed threshold T= 70% [21] for any port, links associated with the port is identified as bottleneck link, accordingly the bottleneck links list will be updated in every 10 seconds. Algorithm 1 computes port utilization and identifies bottleneck link as illustrated below.

**Algorithm 1** Identifying Bottleneck Link list

**Input: Threshold T, time-interval t, port Speed**
**Output: List of bottleneck ports**
**Procedure: BottleneckIdentification()**

1. Initialize BottlenecklinkList BL as Empty
2. **for <every switch> do**
3. **for <each port> do** ▷*Compute Port utilization*
4. $U = (tx_Bytes * 8 * 100)/(portSpeed * t)$
5. **If** $U \geq T$ **then**
6. Add link to BottlenecklinkList BL.
7. **end if**
8. **end for**
9. **end for**

### 3.2.2 Alternate path computation

Once bottleneck links are identified the controller invokes re-routing module to compute alternate path excluding the identified bottleneck links. For alternate route computation, network's virtual topology is retrieved and updated locally by setting the weight for bottleneck links as infinity; alternate path computation is performed on this modified network topology without considering the bottleneck links. As illustrated in [26], once the topology is updated, alternate path computation is performed between source and destination pair using Dijikstra's algorithm. The algorithm for alternate path computation is illustrated below in the Algorithm 2.

---

**Algorithm 2** Algorithm to compute Alternate Path

**Input: Bottleneck Link List BL, Impassable List IL, Network Topology**
**Output: Alternate path for bottleneck links**
**Procedure: Alternate Path (BL or IL)**

1. **if** L ≠empty **then**
2. **for** <every link $L_j$ in $L$> **do**
3. Set Topology.$L_j$.weight = infinity
4. State Select the largest flow through BottleneckLink, $L_j$
5. Set src_node = source node of flow
6. Set dest_node = destination node of flow
7. $AltPath_{Lj}$ = Dijikstra's Algorithm (*src_node, dest_node*) ⊳ *Compute the next shortest path for the flow*
8. **end for**
9. **end if**

---

If this link is on the unique path between the nodes, the algorithm will not return any alternate path as this link's weight is made infinity. Then all flows have to take this congested path as there is no alternate path. The OpenFlow protocol maintains *port_stat* and *flow_stat* structures, the fields in these structures contain various flow level information [5]. In [36], flow level statistics are computed through periodic monitoring and collection of flow level information maintained in these fields. The largest flow is selected based on the volume of data transmitted for the flow. Byte count field in *flow_ stat* structure gives the number of bytes in a flow entry which in turn is used to select the largest flow through the link as illustrated in [34].

### 3.2.3 Flow admission module

The flow admission module is responsible for admitting flows to appropriate paths by ensuring the congestion-free transmission of packets from source to destination. As in [23], before transmitting a flow through a particular path, the SDN controller determines whether the links in the path have sufficient bandwidth to accommodate the flow and the link is available for the duration over which the flow is to be transmitted. The residual bandwidth of each link through which the flow has to be sent is calculated. In [17] and [35], the links with sufficient residual bandwidths are the probable links through which the packets can be sent. If the residual bandwidth is available for all the links in the alternate path, the controller calculates the Bayesian network probability of each link to ensure the link available in the alternate path. The following network parameters are used for Bayesian network modeling:

1. Observation Window (OW): The time period over which the port monitoring is performed and the traffic statistics are collected.
2. Port Utilization (PU): The portion of time the ports are being used during an observation window is directly observable from the port status request reply messages. The probability of port utilization *P(PU)* is assigned as the observed port utilization.

3. Residual Bandwidth (RB): The remaining bandwidth of a link after the requested flow is allocated.

$$RB = \mathrm{BW}_{\mathrm{available}} - \mathrm{BW}_{\mathrm{requested}} \tag{2}$$

   Available bandwidth can be computed by knowing the capacity of the link and transmitted bytes counts *tx_bytes* from the flow table. Requested bandwidth can be obtained from the flow request. If *RB* is positive then *P(RB)* is set to 1, otherwise *P(RB)* is set to 0.

4. Link Availability (LA): Decides whether the flow can be admitted through the given link. Based on the residual bandwidth and port utilization over the Observation Window the link availability is computed using the formula.

$$P(LA|RB,PU) = \frac{P(RB,PU|LA)P(LA)}{P(RB)P(PU)} \tag{3}$$

The graphical model of Bayesian Network showing the relationship between the network parameters considered for deciding the flow admission is depicted in Fig. 2.The Bayesian Network model computes the probability of flow admission from the dynamically computed parameters Port Utilization and Residual Bandwidth within a given Observation Window. Flows are re-routed via an alternate path only if it can handle the new flow load without causing further congestion. If the probability of link availability for each link in the alternate path is larger than the probability that it is unavailable, that path is the right candidate to accommodate the flow. The newly computed path is then installed in the flow tables of all the switches that are part of this new path. New flows are installed to flow tables by setting a hard timeout value so that traffic gets routed via the best path. Once the newly computed flow is applied to the flow tables, further traffic uses this newly applied flow until the flow's hard timeout expires, decreasing congestion and improving network efficiency. Using the modelled Bayesian network, Algorithm 3 computes the possibility of flow admission.

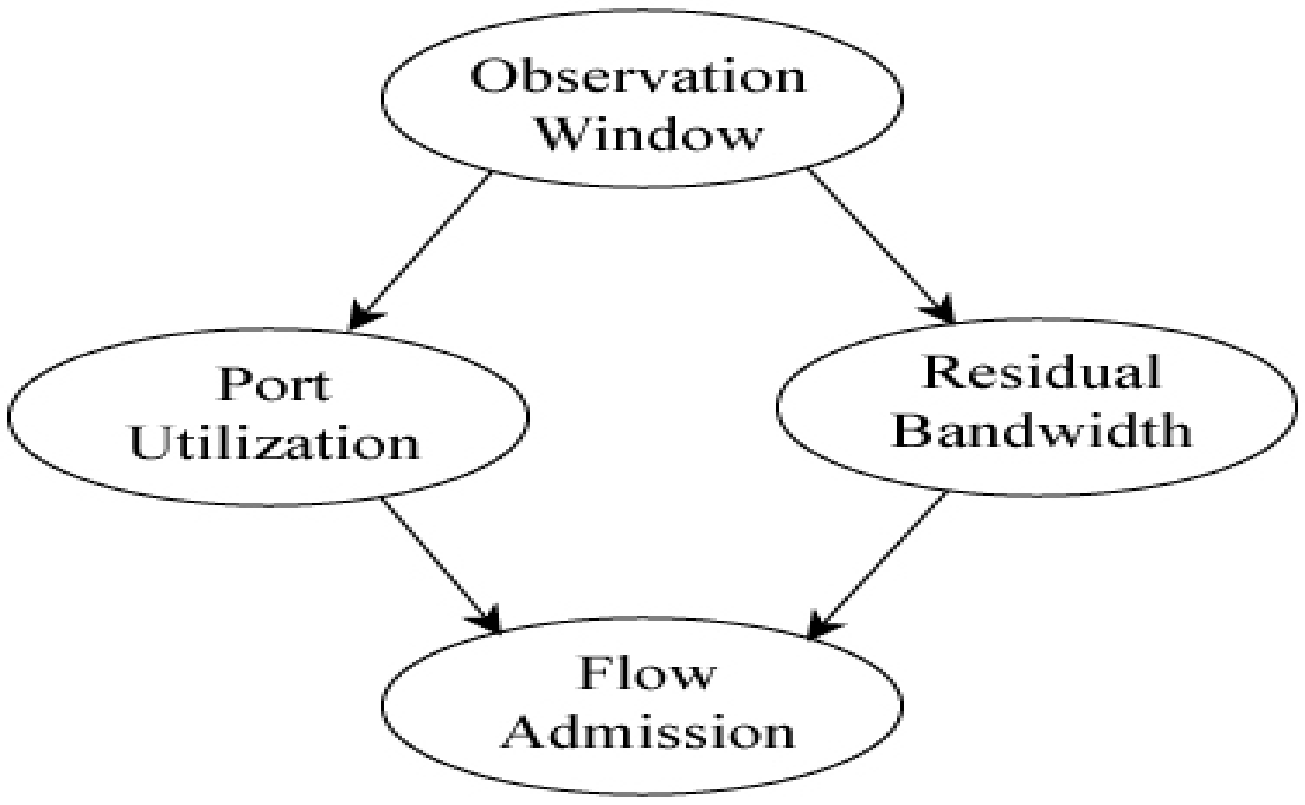

**Fig. 2** Bayesian Network Model For Computing Flow Admission

**Algorithm 3 Algorithm to Compute Flow Admission**

**Input: Link capacity, Flow bandwidth request, Available bandwidth, Observation window, Alternate path**
**Output: ImpassablelinkList, IL**
**Procedure: FlowAdmission()**

Initialize *IL* to empty
  **for** < each link $L_j$ in the AltPath > **do**
    $BW_{available}$ = *Capacity - tx Bytes_Transmitted*
    Residual Bandwidth, RB $= BW_{available} - BW_{requested}$
    **if** $RB > 0$ **then** ▷ Compute the link availability using Bayesian network probabilities
      $P(RB) = 1$
      Compute

$$P(LA|RB,PU) = \frac{P(RB,PU|LA)P(LA)}{P(RB)P(PU)}$$

      **if** $(P(LA) > (1 - P(LA)))$ **then** ▷ Decide whether the link is available
        set $LA=1$ for link $L_j$
      **else**
        set $LA=0$ for link $L_j$
        Add the link to *IL*
      **end if**
  **else** ▷ Remove the links with insufficient bandwidth
        Add the link to *IL*
    **end if**
**end for**
**if** $IL \neq empty$ then
  call Alternate Path (*IL*)
**end if**
**if** L=True $\forall L_j \in$ AltPath **then**
  **for** < every switch $N_i$ in AltPath > **do**
    Add a flow table entry from *src_node* to *dest_node* and vice-versa in switch $N_i$
  **end for**
**end if**

## 4. SIMULATION AND RESULTS

Mininet is used to simulate the Network topology given in Fig. 3 with 11 hosts and switches. Mininet supports both built-in and external SDN controller, switches in Mininet use OpenFlow protocol [11].

In the simulation the external SDN controller is implemented using OpenIris SDN controller based on the Floodlight controller developed by ETRI which is specified in [13]. OpenIris controller based on event handling is written in Java, the three new modules in the proposed system are included as the subclass of OF Module superclass.

The tool, Distributed Internet Traffic Generator (D-ITG) is used to generate IPv4 and IPv6 traffic at packet level using appropriate stochastic process. D-ITG is capable of measuring different network performance metrics by analyzing the log files generated by ITGSend and ITGRecv modules [12]. In the simulation of proposed system, D-ITG is configured to follow exponential distribution for both IDT (Inter Departure Time) and PS (Packet Size).

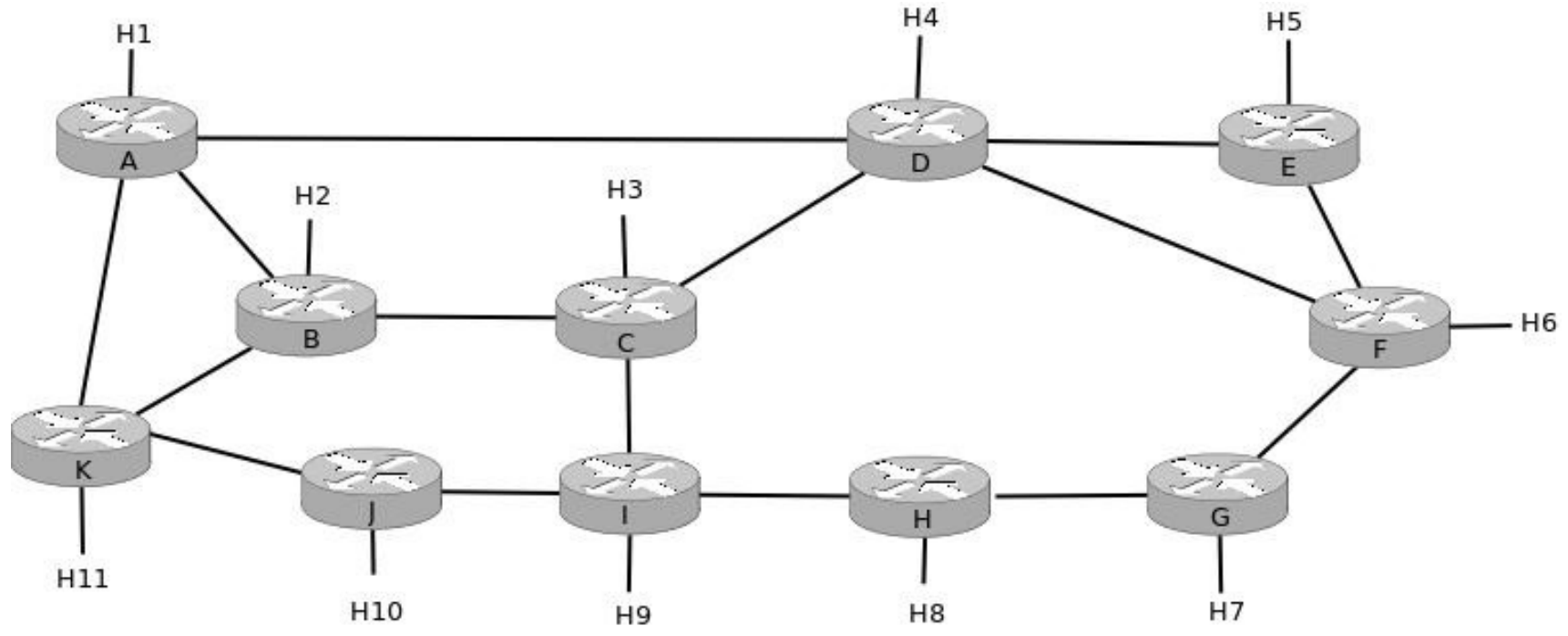

**Fig. 3** Network topology

The network topology for simulation is Abilene network taken from the standard topology available in the Internet Topology Zoo [27] and [32].

## 4.1 Performance Analysis

The performance of the proposed system is analyzed in terms of average packet loss, average throughput and average end-to-end delay [14] and [25]. The above performance metrics are measured for the same topology by varying the number of flows in the network. Let N be the total number of flows.

*Packet Loss*: Packet loss is the number of packets that fails to arrive at the destination. It is represented as loss percentage. Packet loss for each flow is calculated using the formula:

$$Packet\ Loss = \frac{Number\ of\ packets\ dropped}{Total\ Number\ of\ packets\ send} * 100 \tag{4}$$

*Average packet loss*: is the percentage packet loss averaged over N flows calculated using the formula:

$$Average\ Packet\ Loss = \frac{\sum_{i=1}^{N} Packet\ loss\ of\ each\ flow\ i}{N} \tag{5}$$

*Average Throughput*: It is the average amount of data delivered in unit time represented in Mbps. Average throughput is calculated using the formula:

$$Average\ Throughput = \frac{\sum_{i=1}^{P} \text{Data bits received in flow i}}{\text{Total simulation time} * N} \tag{6}$$

*Average end-to-end delay*: The end-to-end delay of a flow is the time taken by all packets to reach destination averaged over all the P packets transmitted. It is calculated using the formula:

$$End\text{-}to\text{-}end\ Delay = \frac{\sum_{i=1}^{P} PacketEndTime\text{-}PacketStartTime}{P} \tag{7}$$

Average end-to-end delay for the network is the end-to-end delay averaged over total number of flows.

$$\text{Average end-to-end Delay} = \frac{\sum_{i=1}^{N} \text{end-to-end Delay of Each Flow } i}{N} \quad (8)$$

The proposed system evaluation is done in two simulation scenarios by varying packet rate and varying the number of flows.

### 4.1.1 Single flow with varying packet rate

In the first scenario, link AB in the above network topology is configured at 10 Mbps bandwidth and all other links at 20 Mbps bandwidth. Using D-ITG packet generator, 1000 bytes packets are sent from host H1 to host H2 by varying the number of packets transmitted per second. The performance of the proposed system is analyzed in a single flow scenario in terms percentage of packet loss and throughput and compared the results with the existing network without rerouting.

#### 4.1.1.1. Packet Loss

The observed percentage of packet loss with varying packet rate in the first simulation scenario is as shown in Table 1.

Table1. Percentage Packet loss with varying packet rate

| Number of packets send(pps) | Without Re-routing (packet loss in %) | With Re-routing (packet loss in %) |
|---|---|---|
| 1000 | 0 | 0 |
| 2000 | 17.93 | 4.90 |
| 3000 | 38.05 | 9.50 |
| 4000 | 48.97 | 13.91 |
| 5000 | 53.43 | 16.20 |
| 10000 | 74.03 | 56.05 |

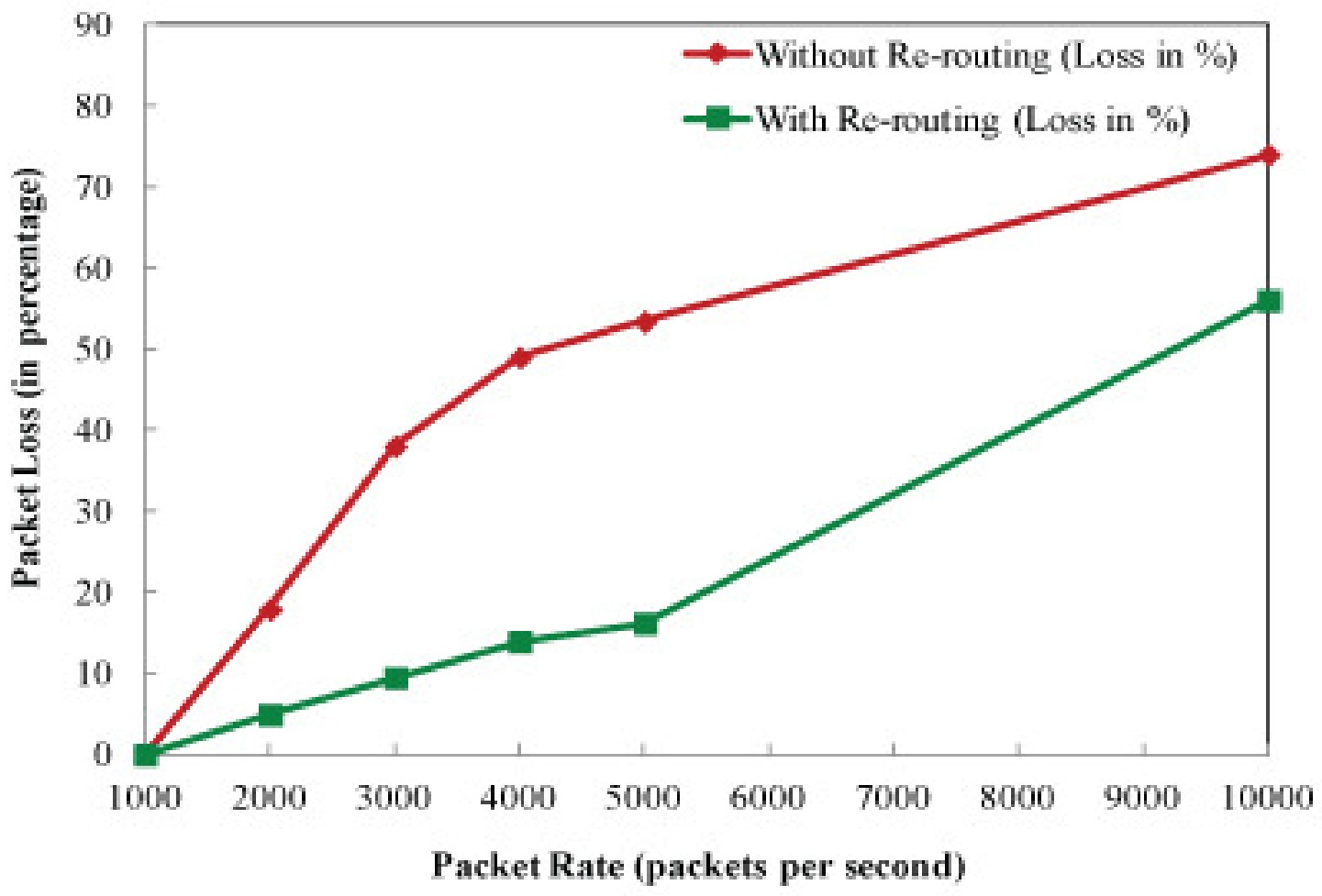

**Fig. 4** Percentage Packet loss with varying packet rate

It is observed that the proposed system provides lesser packet loss than an existing system without rerouting. Proactively re-routing the packets via the alternate path, the proposed system considerably reduces the number of packets dropped. The percentage of packet loss with varying packet rate is as shown in Figure 4.

#### 4.1.1.2. Throughput with Varying Packet Rate

Observed throughput with varying packet rate in the first simulation scenario is as shown in Table 2.

Table 2. Throughput with varying Packet Rate

| Number of packets send(pps) | Without Re-routing (Throughput in Mbps) | With Re-routing (Throughput in Mbps) |
|---|---|---|
| 1000 | 6.52 | 6.51 |
| 2000 | 9.69 | 11.06 |
| 3000 | 9.70 | 13.99 |
| 4000 | 9.70 | 16.15 |
| 5000 | 9.69 | 16.57 |
| 10000 | 9.70 | 16.62 |

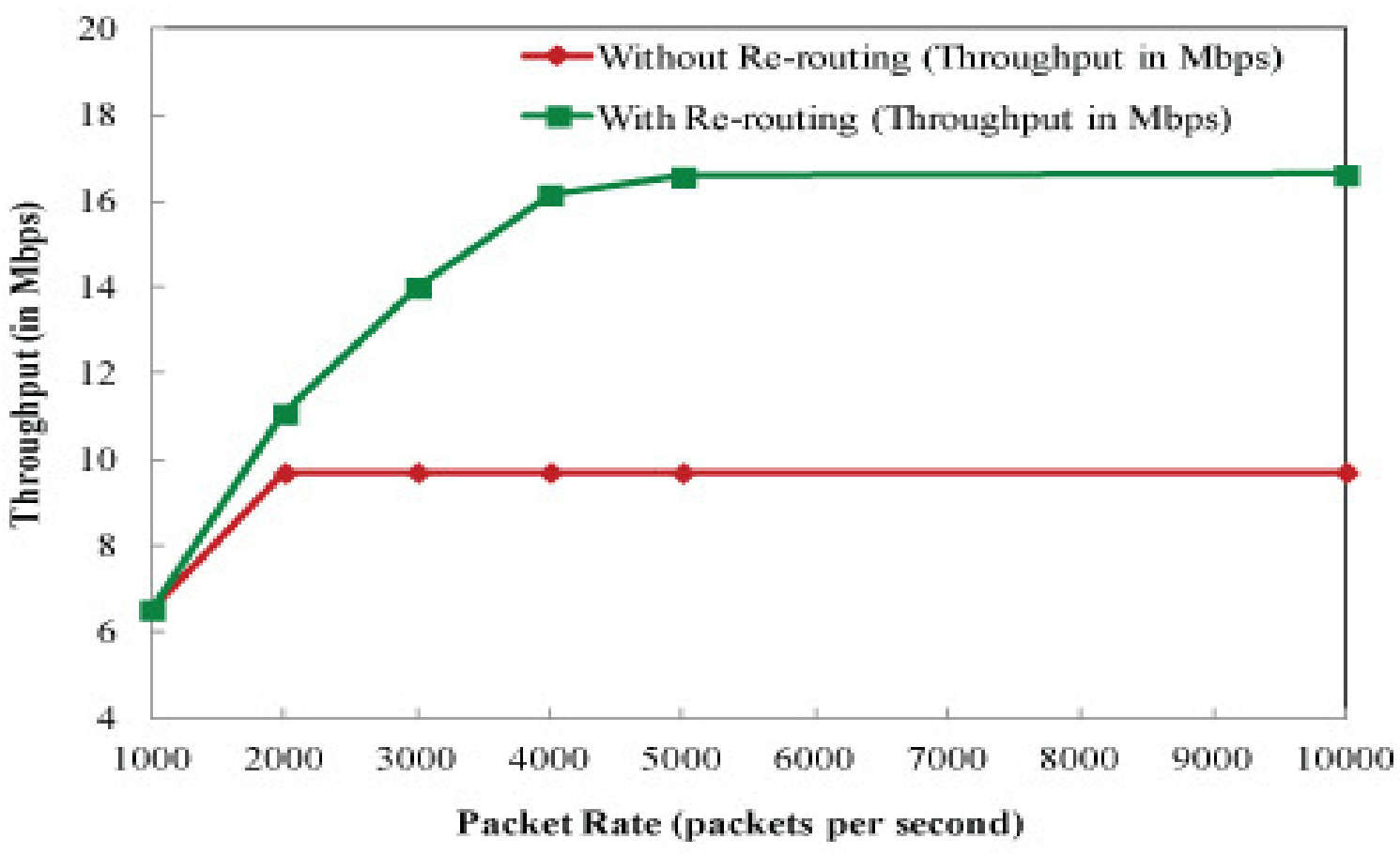

**Fig. 5** Throughput with varying Packet Rate

In the simulation scenario bottleneck link capacity is fixed at 10 Mbps hence the throughput increases with packet rate then remained approximately same at 9.7 Mbps irrespective of the variation in packet rate for the existing system without rerouting. It is observed that throughput of the proposed system is better than the existing system, by using the alternate path for re-routing, the proposed system reroutes packets through the alternate path without bottleneck links hence reducing packet loss and achieving higher throughput up to 16.62 Mbps since all other links are rate limited to 20 Mbps. The throughput with varying packet rate is depicted in Fig 5.

### 4.1.2 Multiple flows with a fixed packet rate

In the second scenario, each port in each switch is rate limited at 20 Mbps. This is done by creating the queue in each port and configuring its max rate as 20 Mbps. 1000 bytes packets are then sent from host *H1* to all other host for 60 seconds at 1000 packets per second. So 8 Mbps traffic is sent from *H1* to all other hosts in the test topology amounting to a maximum of 10 flows in the network. Simulation is conducted to compare the performance of ARLD reactive re-routing approach used in [3] and the proposed proactive re-routing strategy. The simulation results obtained for the QoS parameters mentioned above is shown below.

#### 4.1.2.1 Average Packet Loss

The average packet loss with varying number of flows is given in Table 3.

Table 3. Average Packet Loss

| Number of flows | Reactive Re-routing (Loss in %) | Proactive Re-routing (Loss in %) |
|---|---|---|
| 1 | 0 | 0 |
| 2 | 0 | 0 |
| 3 | 0 | 0 |
| 4 | 0 | 0 |
| 5 | 2.50 | 2.39 |
| 6 | 2.91 | 2.63 |
| 7 | 3.70 | 3.45 |
| 8 | 15.55 | 4.49 |
| 9 | 25.69 | 15.09 |
| 10 | 33.16 | 23.78 |

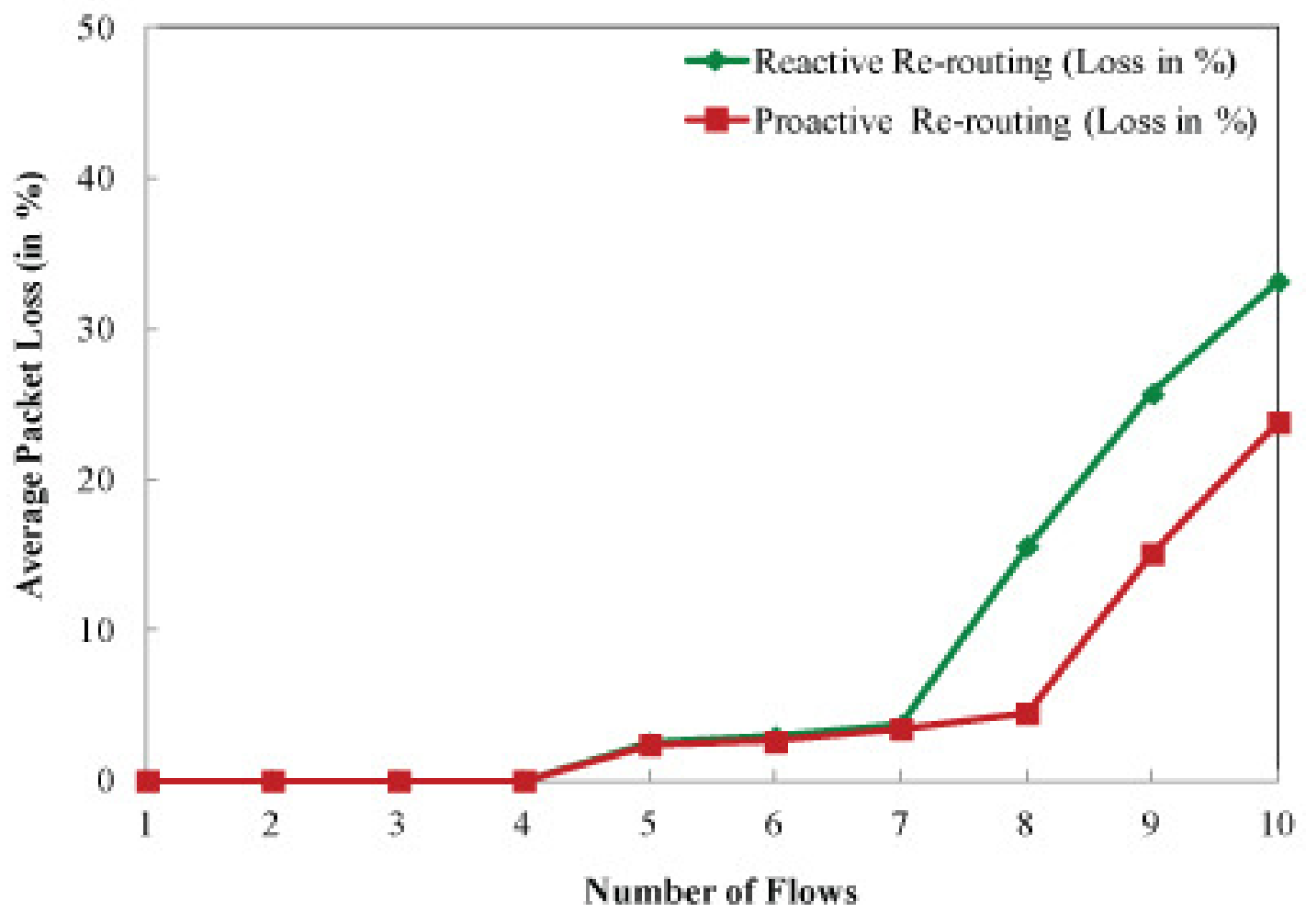

**Fig. 6** Average Packet Loss

It is observed that the proposed system with proactive re-routing reported much lesser average packet loss than the system with reactive re-routing. Initially, packet loss remained the same for both the system as the network was not congested. When the number of flows increased, proactive re-routing algorithm results in much lesser packet loss than reactive re-routing scheme. Fig. 6 shows the variation of average packet loss with varying number of flows.

#### 4.1.2.2 Average Throughput

The average throughput with varying number of flows is as shown in Table 4.

Table 4. Average Throughput

| Number of flows | Reactive Re-routing (Mbps) | Proactive Re-routing (Mbps) |
|---|---|---|
| 1 | 6.55 | 6.54 |
| 2 | 6.58 | 6.58 |
| 3 | 6.81 | 6.80 |
| 4 | 6.83 | 6.83 |
| 5 | 6.54 | 6.53 |
| 6 | 6.56 | 6.56 |
| 7 | 6.43 | 6.43 |
| 8 | 5.64 | 6.46 |
| 9 | 5.02 | 5.74 |
| 10 | 4.51 | 5.18 |

It is observed that proactive re-routing reported much better throughput than reactive re-routing. By proactively rerouting network traffic, packet drop is reduced which resulted in higher throughput. Fig. 7 depicts the variation of average throughput with varying number of flows.

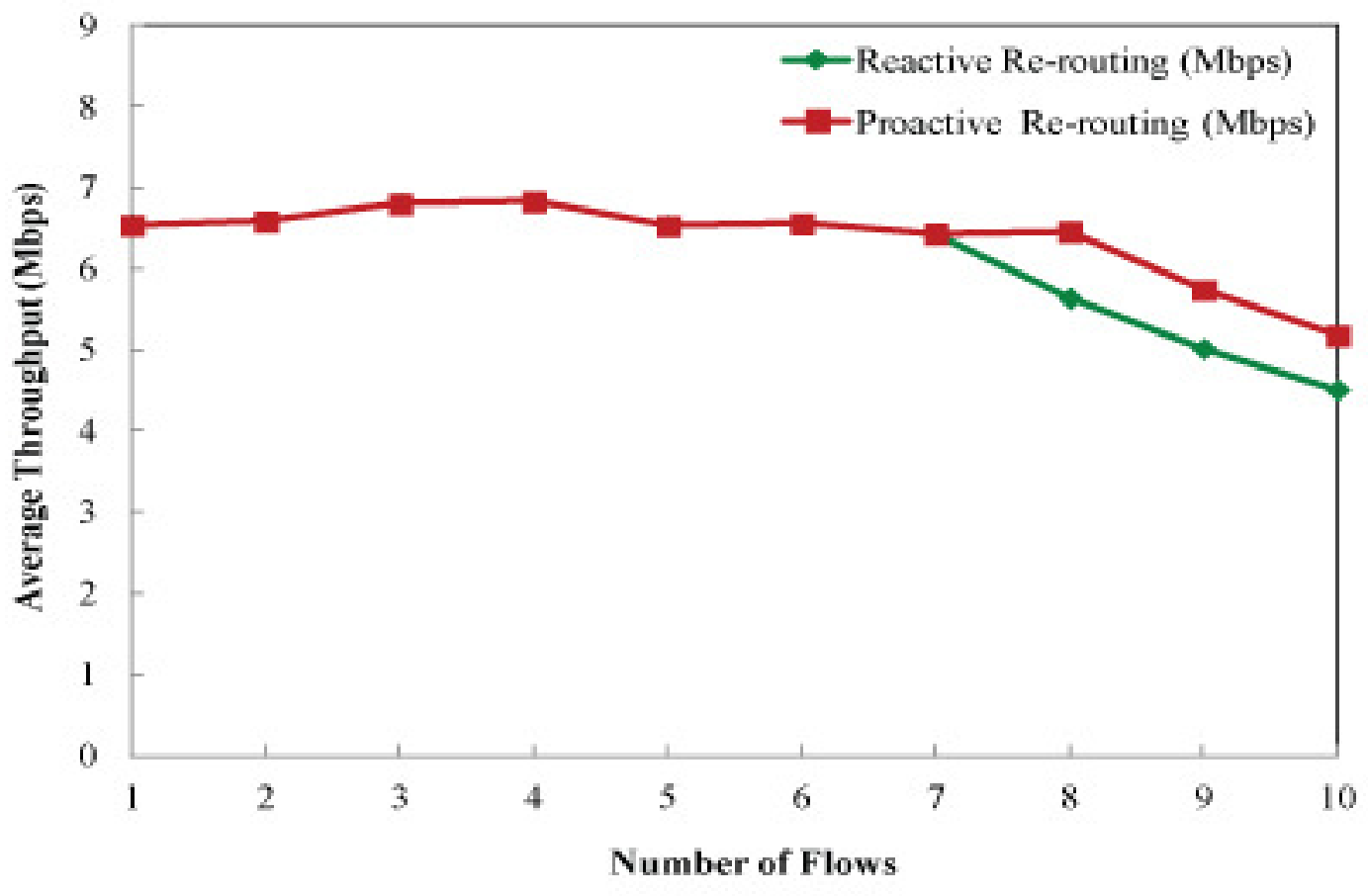

**Fig. 7** Average Throughput

### 4.1.2.3. Average end-to-end Delay

The average end-to-end delay by varying the number of flows is as shown in Table 5.

Table 5. Average end-to-end Delay

| Number of flows (pps) | Reactive Re-routing(ms) | Proactive Re-routing(ms) |
|---|---|---|
| 1 | 0.043 | 0.042 |
| 2 | 0.046 | 0.045 |
| 3 | 0.049 | 0.047 |
| 4 | 0.055 | 0.050 |
| 5 | 0.050 | 0.050 |
| 6 | 233.596 | 233.596 |
| 7 | 312.797 | 305.536 |
| 8 | 321.991 | 324.633 |
| 9 | 354.49 | 332.586 |
| 10 | 373.629 | 335.569 |

From the Fig. 8, it is observed that the delay incurred by packets is much lesser in the proposed system than in the existing system. Proposed system pro-actively re-routes the packets via less congested links thus delivering packet faster than existing system.

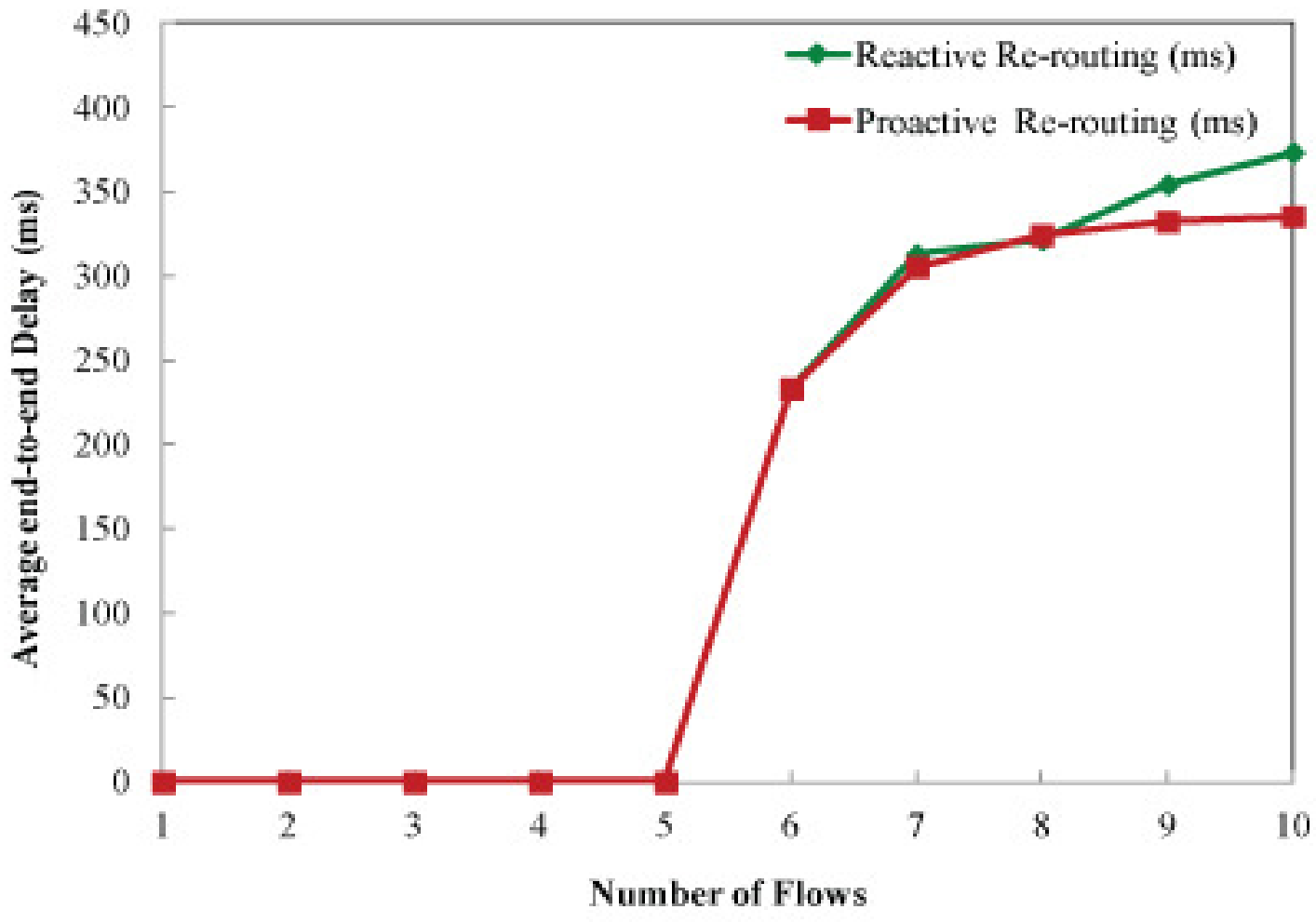

**Fig. 8** Average end-to-end Delay

The simulation results show that in both experiment scenarios when the network load escalates, the proposed method re-routes the network traffic which results in efficient load balancing. The use of Bayesian Network model to decide flow admission through the newly computed alternate path mitigates the possibility of congestion propagation. Hence from the perspective of performance metrics, it is seen that the proposed method improves the average network throughput and reduces the average packet loss and end-end delay compared to existing load balancing schemes in SDN.

## 5. Conclusions and Future Work

The proposed algorithms implement a proactive traffic analysis based load balancing method which reduces network congestion and congestion propagation in SDN. Through periodic monitoring of flow-based statistics, the proposed method re-routes network traffic from a congested path to a non-congested alternate path. A Bayesian network is modelled to ascertain that the flow rerouting through the alternate path computed will not lead to further congestion and avoids congestion propagation as well. The simulation results show that the proposed proactive traffic analysis method to reduce congestion and packet loss outperformed existing network without re-routing in terms of packet loss and throughput as well as the reactive method from the perspective of average packet loss, average throughput and average end-to-end delay. However, periodic flow monitoring based on query-response for traffic analysis used in the proposed system introduces an additional overhead to the controller. As a future work multi-controller scenario using standard network topologies can be implemented to reduce the SDN controller overhead.

## AUTHORS

Sminesh C. N. received his ME degree from Birla Institute of Technology, Ranchi, India in the year 2003. He is working as Associate Professor in the Department of Computer Science and Engineering, Government Engineering College Thrissur. His research interests include Traffic Engineering in SDN, Next Generation Internet Architecture. He is the author of more than 15 research articles published in national and international conferences

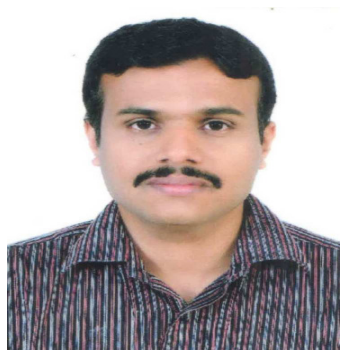

Grace Mary Kanaga E. received her PhD in Computer Science and Engineering from Anna University Coimbatore, India in the year 2011. She is working as an Associate Professor in Computer Sciences Technology Department, Karunya Institute of Technology and Sciences, Coimbatore. Her research interests are Big Data Analytics, Computational Intelligence, Software Agents, Networks and Distributed Systems. She has published more than 45 papers in National and International Conferences, International Journals and Book chapters.

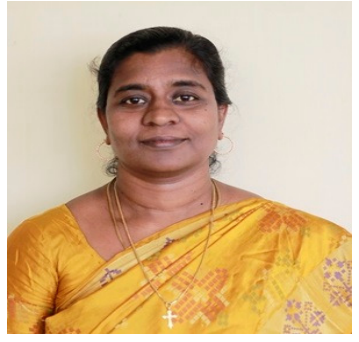

Ranjitha K. received her ME degree from Government Engineering College Thrissur, India in the year 2016. Presently she is a Senior Research Fellow, CR Rao AIMSCS, University of Hydearbad Campus, Hyderabad, India. Her research interests include Information Security and Future Internet. She has seven years of Industry experience and author of several research articles.

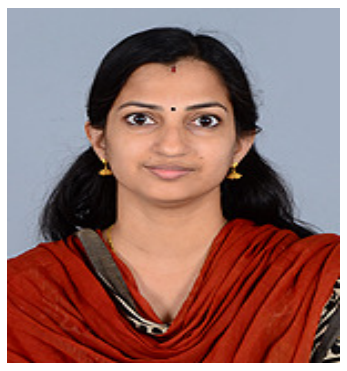